\documentstyle[epsfig,eqsecnum,aps]{revtex}
\begin{document}
\draft
\title{Half-lives of rp-process waiting point nuclei}
\author{P. Sarriguren, R. \'Alvarez-Rodr\'{\i}guez, E. Moya de Guerra}

\address{Instituto de Estructura de la Materia,
Consejo Superior de Investigaciones Cient\'{\i }ficas, \\
Serrano 123, E-28006 Madrid, Spain}
\maketitle

\begin{abstract}

We give results of microscopic calculations for the half-lives of 
various proton-rich nuclei in the mass region A=60-90, which are 
involved in the astrophysical rp-process, and which are needed 
as input parameters of numerical simulations in Nuclear Astrophysics.
The microscopic formalism consists of a deformed QRPA approach that 
involves a selfconsistent quasiparticle deformed Skyrme Hartree-Fock 
basis and residual spin-isospin separable forces in both the 
particle-hole and particle-particle channels. The strength of the
particle-hole residual interaction is chosen to be consistent with
the Skyrme effective force and mean field basis, while that
of the particle-particle is globally fixed to $\kappa =0.07$ MeV
after a judicious choice from comparison to experimental half-lives.
We study and discuss the sensitivity of the half-lives to deformation 
and residual interactions.

\end{abstract}


\section{Introduction}
\label{intro}

Many problems in Nuclear Astrophysics usually require the use of numerical 
simulations and network calculations using weak interaction rates as input 
parameters \cite{grotz}.
High quality nuclear input is a necessary condition for a high quality
astrophysical model. Thus, the models for the energy generation in stars 
and for nucleosynthesis may depend critically on the nuclear input used.

In particular, the rapid proton capture (rp) process is of special interest 
\cite{schatz}. It is expected to take place in explosive scenarios, such as 
x-ray bursts, where the necessary conditions of high densities and 
temperatures are met. Actually, the relevance of the nuclear rp-process 
network lies in its importance as the dynamical engine of the observed 
x-ray bursts. The rp-process is characterized by the fact that the proton 
capture reaction rates are orders of magnitude faster than the competing 
$\beta^+$ decays. When the proton capture is inhibited, the reaction flow 
has to wait for a relatively slow $\beta^+$ decay to continue \cite{schatz}. 
The longer lived $\beta^+$ emitters are called waiting points. The half-lives 
of the waiting point nuclei determine the time scales of the flow.

Calculations of $\beta^+$/EC half-lives have been carried out in the past
at different levels of approximation. The first large scale calculations
were based on the gross theory \cite{taka}. This is a statistical model
that averages over the $\beta$-strength distributions in the daughter 
nucleus ignoring the intrinsic nuclear structure. 

More recently, many efforts have been done to calculate the weak 
interaction rates from microscopic calculations taking into account the 
nuclear structure details of the individual nuclei, see for 
instance \cite{moellernk,nabi,langanke}. Among the microscopic methods, 
the proton-neutron quasiparticle random phase approximation (pnQRPA or 
QRPA) is one of the most reliable and widely used microscopic 
approximations for calculating the correlated wave functions involved 
in $\beta$ decay processes. The method was first studied in 
Ref. \cite{halb} to describe the $\beta$ strength distributions. The 
inclusion of a particle-particle ($pp$) residual interaction \cite{vogel}, 
in addition to the particle-hole ($ph$) usual channel, and the extension 
of the method to deal with deformed nuclei \cite{kru,moll,homma} using 
phenomenological potentials, were major steps to improve the method.

Selfconsistent methods have been also applied to the study of the decay
properties of spherical neutron-rich nuclei \cite{doba} and deformed
proton-rich nuclei \cite{sarri1,sarri2,sarri3}. In the latter work, 
$\beta $-decay 
properties were studied on the basis of a deformed 
HF+BCS+QRPA calculation with density dependent effective interactions 
of Skyrme type. This is indeed a very suitable method to study the 
$\beta$-decay half-lives of the exotic proton-rich waiting point nuclei 
involved in the rp-process, namely the $N=Z$ nuclei $^{64}$Ge, $^{68}$Se,
$^{72}$Kr, $^{76}$Sr, $^{80}$Zr, $^{84}$Mo, $^{88}$Ru, and $^{92}$Pd.
The experimental information on half-lives existing on these nuclei 
is used to test the model results and to asses the predictive power 
of the method for network calculations of the rp-process.

The deformed HF+BCS+QRPA with density-dependent Skyrme forces and residual
interactions consistent with the mean field is a method that has been 
successfully used in the description of the nuclear structure properties 
of nuclei within the valley of stability. The quality of the two-body 
effective Skyrme interaction and the selfconsistent procedure will 
finally determine the extrapolability to unknown exotic regions. Nuclear 
deformation is a relevant ingredient in this mass region \cite{moeller}, 
which is crossed by the rp-path, and should be included in a reliable 
calculation. The procedure determines the deformation selfconsistently, 
which is again a clear advantage in regions where there is no experimental 
information on nuclear shapes. Following this method, we study in this 
work the half-lives of the above mentioned waiting points and their 
dependence on deformation and residual forces.

The paper is organized as follows. In Section 2, we present a brief summary 
containing the basic points in our theoretical description. Section 3 
contains the results obtained for the ground state properties of the $N=Z$ 
waiting point nuclei considered, as well as the half-lives and their 
dependence on deformation and residual forces. The conclusions are given in 
Section 4.

\section{Brief summary of the theoretical formalism}
\label{sec:2}

In this Section we summarize briefly the theoretical formalism used to 
describe the Gamow-Teller transitions. More details can be found in 
Ref. \cite{sarri1,sarri2,sarri3}.

We follow a selfconsistent Hartree-Fock procedure to generate 
microscopically the deformed mean field, which is assumed to be axially 
symmetric. This is done with density dependent effective interactions of 
Skyrme type. The equilibrium deformation of the nucleus is obtained 
selfconsistently as the shape that minimizes the energy of the nucleus.
In this work we present results obtained with the most traditional of
the Skyrme forces (Sk3) \cite{beiner} and with the force SG2 \cite{giai},
which is known to provide a good description of the spin-isospin excitations
in nuclei.

The single-particle wave functions are expanded in terms of the eigenstates 
of an axially symmetric harmonic oscillator in cylindrical coordinates, 
using eleven major shells in the expansion. Pairing correlations between 
like nucleons are included in the BCS approximation with fixed gap 
parameters for protons and neutrons. The gap parameters are determined 
phenomenologically from the odd-even mass differences.

Following Bertsch and Tsai \cite{btsai}, the particle-hole interaction
consistent with the Hartree-Fock mean field is obtained as the second
derivative of the energy density functional with respect to the one-body
density. Neglecting momentum dependent terms, the resulting local 
interaction can be written in the Landau-Migdal form. After functional
differentiation, one can establish a relationship between the Landau
and the Skyrme parameters. The resulting residual local force is now
approximated \cite{sarri1} by a separable force by averaging the local 
force over the nuclear volume assuming a constant density distribution 
inside a sphere with the nuclear radius. Integrating over this volume, 
we arrive to a separable spin-isospin force whose coupling strength is 
determined by the Skyrme parameters and the equilibrium radius. Hence, 
it varies in accordance with the Skyrme force used. 
The reliability of this procedure was discussed in Refs. \cite{m1,zawisha}
in the context of the spin magnetic dipole excitations, which are the 
$\Delta T_z=0$ isospin counterparts of the $\Delta T_z =\pm 1$ GT transitions.
The conclusion was that, although less realistic, the separable interaction
contains the essential features of the zero-range force \cite{zawisha}.
In summary, the method is a compromise between exact consistency and
manageability of the residual force, which is a separable residual interaction
whose strength is consistent with the ground state mean field.
This procedure can be viewed as an approximation to the more general
method \cite{giai2}, where the exact $ph$ residual interaction is first 
reduced to its Landau-Migdal form and then the RPA matrix is expanded
into a finite sum of $n$ separable terms. One may expect that for radial
independent RPA modes, which are considered here, the approximation 
made here gives a good average of the above mentioned expansion. The results 
in Ref. \cite{zawisha} support this view.

The particle-particle part is a neutron-proton pairing force in the 
$J^{\pi}=1^+$ coupling channel. We introduce this interaction in terms
of a separable force with a coupling strength, $\kappa ^{pp}_{GT}$, 
determined by fitting the $\beta$-decay half-lives of $\beta$ emitters
at their equilibrium shapes.

The basic quantities needed to 
calculate the $\beta$-decay properties are the matrix elements connecting 
proton and neutron states with Fermi or Gamow-Teller operators.

We introduce the proton-neutron QRPA phonon operator for GT excitations 
in even-even nuclei

\begin{equation}
\Gamma _{\omega _{K}}^{+}=\sum_{\pi\nu}\left[ X_{\pi\nu}^{\omega _{K}}
\alpha _{\nu}^{+}\alpha _{\bar{\pi}}^{+}+Y_{\pi\nu}^{\omega _{K}}
\alpha _{\bar{\nu}} \alpha _{\pi}\right]\, ,  \label{phon}
\end{equation}
where $\alpha ^{+}\left( \alpha \right) $ are quasiparticle creation
(annihilation) operators, $\omega _{K}$ are the RPA excitation energies, 
and $X_{\pi\nu}^{\omega _{K}},Y_{\pi\nu}^{\omega _{K}}$ the forward and 
backward amplitudes, respectively. 

In the intrinsic frame the GT transition amplitudes connecting the QRPA 
ground state $\left| 0\right\rangle$ to one phonon states 
$\left| \omega _K \right\rangle$ satisfying

\begin{equation}
\Gamma _{\omega _{K}} \left| 0 \right\rangle =0\, , \qquad
\Gamma ^+ _{\omega _{K}} \left| 0 \right\rangle = \left| 
\omega _K \right\rangle \, ,
\end{equation}
are given by

\begin{equation}
\left\langle \omega _K | \sigma _K t^{\pm} | 0 \right\rangle = 
\mp M^{\omega _K}_\pm \, ,
\end{equation}
where

\begin{eqnarray}
M_{-}^{\omega _{K}}&=&\sum_{\pi\nu}\left( q_{\pi\nu}
X_{\pi\nu}^{\omega _{K}}+ \tilde{q}_{\pi\nu}
Y_{\pi\nu}^{\omega _{K}}\right) \, , \nonumber \\
M_{+}^{\omega _{K}}&=&\sum_{\pi\nu}\left( \tilde{q}_{\pi\nu}
X_{\pi\nu}^{\omega _{K}}+
q_{\pi\nu}Y_{\pi\nu}^{\omega _{K}}\right) \, ,
\end{eqnarray}
and
\begin{equation}
\tilde{q}_{\pi\nu}=u_{\nu}v_{\pi}\Sigma _{K}^{\nu\pi },\quad  
q_{\pi\nu}=v_{\nu}u_{\pi}\Sigma _{K}^{\nu\pi}\, ,
\end{equation}
\begin{equation}
\Sigma _{K}^{\nu\pi}=\left\langle \nu\left| \sigma _{K}\right| 
\pi\right\rangle \, ,
\label{qs}
\end{equation}
where $v'$s are occupation amplitudes ($u^2=1-v^2$).

The GT strength $B(GT^\pm)$ in the laboratory system for a transition 
$I_iK_i (0^+0) \rightarrow I_fK_f (1^+K)$ can be obtained as
\begin{equation}
B(GT^\pm ) = \left[ \delta_{K,0}\left\langle 
\omega_{K} \left| \sigma_0t^\pm \right| 0 \right\rangle ^2 +
2\delta_{K,1} \left\langle \omega_{K} \left| \sigma_1t^\pm \right| 
0 \right\rangle ^2 \right] 
\label{bgt}
\end{equation}
in $[g_A^2/4\pi]$ units. 
We have used the initial and final states in the laboratory frame 
expressed in terms of the intrinsic states using the Bohr-Mottelson
factorization \cite{b-m}. The effect of angular momentum projection is 
then, to a large extent, taken into account. In the deformed cases,
the spurious contributions to the strengths coming from higher angular
momentum components in the wave functions are of the order
$<J_{\perp}^2>^{-2}$ \cite{moya}, where $<J_{\perp}^2>$ is the 
ground state expectation value of the angular momentum operator
component perpendicular to the symmetry axis. Values of $<J_{\perp}^2>$
for the deformed nuclei considered in this work are always much larger
than 10 \cite{sarri1} and therefore, spurious contributions are expected to 
represent less than $1\%$ effect.

Similarly, the Fermi strength is obtained as

\begin{equation}
B(F^\pm ) = \left| \left\langle \omega \left| t^\pm \right| 0 
\right\rangle \right|^2 \, ,
\label{bf}
\end{equation}
in $[g_V^2/4\pi]$ units.

The $\beta$-decay half-life is obtained by summing up all the allowed 
transition probabilities weighted with some phase space factors up to 
states in the daughter nucleus with excitation energies lying within 
the $Q$-window,   

\begin{eqnarray}
T_{1/2}^{-1}&=&\frac{1}{D}\sum_{\omega }f\left( Z,\omega \right)
B_{\omega}\, , \nonumber \\
B_{\omega}&=&B_{\omega}(F)+ A^2 B_{\omega}(GT)\, ,
 \label{t12}
\end{eqnarray}
where $f\left( Z,\omega \right) $ is the Fermi integral \cite{gove} and 
$D=6200$~s. We include standard effective factors

\begin{equation}
A^{2}=\left[ \left( g_{A}/g_{V}\right) _{\rm eff}\right] ^{2}=\left[
0.75\left( g_{A}/g_{V}\right) _{\rm free}\right] ^{2} \, . \label{quen}
\end{equation}
In $\beta^+/EC$ decay, the Fermi function consists of two parts, positron 
emission and electron capture. In this work we have computed them 
numerically for each value of the energy, as explained in Ref. \cite{gove}.

Although the $\beta^+/EC$ half-lives are dominated by GT transitions, 
Fermi transitions and especially superallowed transitions to the isobaric 
analog state become important for nuclei with $Z>N$. We have calculated 
these contributions and have found that they contribute to a few percent 
reduction effect on the half-lives. This effect is included in the 
numerical results.

\section{Results for half-lives}
\label{sec:3}

In this Section we present first the results for the bulk properties of 
the nuclei under study based on the quasiparticle mean field description. 
We study first the energy surfaces as a function of deformation. For this 
purpose we perform constrained calculations \cite{constraint}, minimizing 
the HF energy under the constraint of keeping fixed the nuclear deformation. 
We can see in Fig. 1 the total HF energy plotted versus the quadrupole 
deformation parameter
\begin{equation}
\beta = \sqrt{\frac{\pi}{5}} \frac{Q_p}{Zr_c^2} \, ,
\label{betadef}
\end{equation}
defined in terms of the microscopically calculated quadru\-po\-le moment 
$Q_p$ and charge root mean square radius $r_c$. 

\begin{center}
\begin{figure}
\resizebox{0.75\textwidth}{!}{\includegraphics{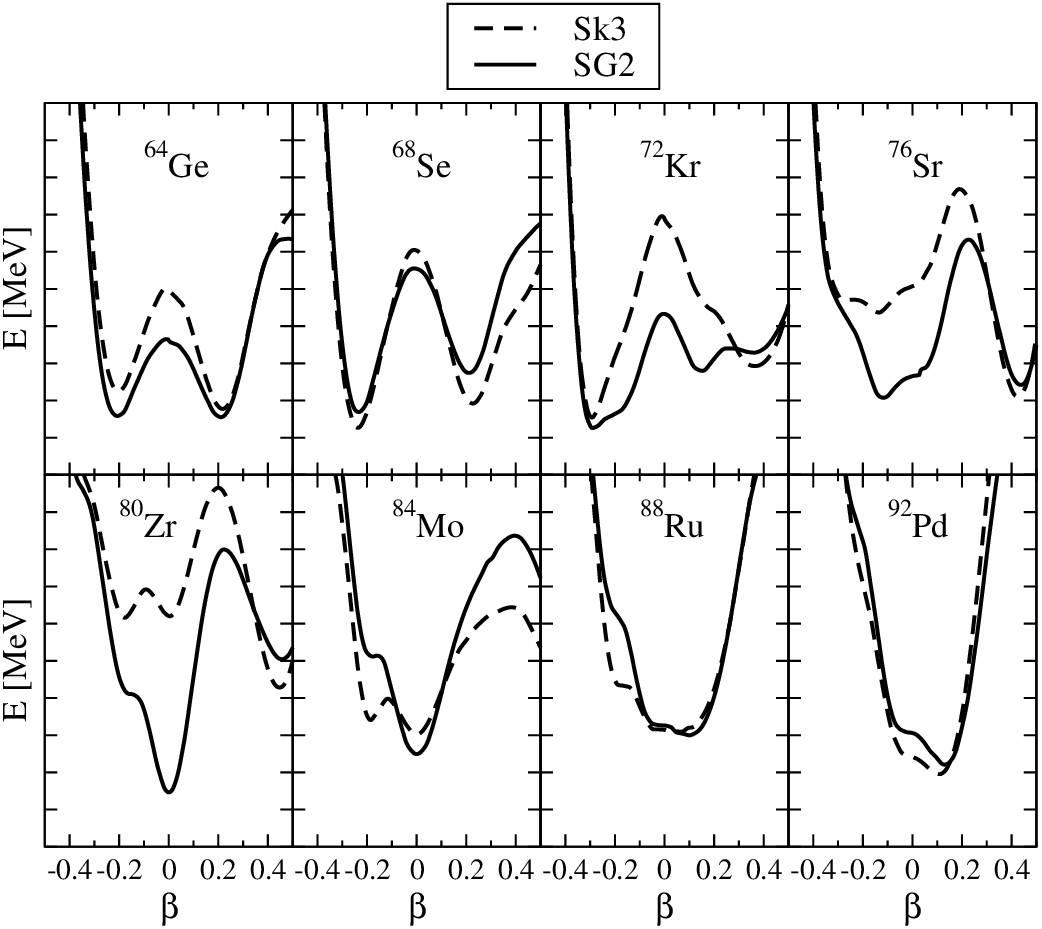}}
\vskip 0.5cm
\caption{Total energy as a function of the quadrupole deformation $\beta$,
obtained from constrained HF+BCS calculations with various Skyrme forces.
The distance between ticks in the vertical axis is always 1 MeV, but the
origin is different in each case.}
\label{fig:1}   
\end{figure}
\end{center}

The results in Fig. 1 correspond to HF+BCS calculations with the forces 
Sk3 and SG2. We observe that both forces predict in most cases 
quite similar equilibrium shapes and structure of the energy curves.
In some instances ($^{64}$Ge, $^{68}$Se, $^{72}$Kr, $^{76}$Sr and 
$^{80}$Zr) we obtain two energy minima close in energy, indicating the 
existence of shape isomers in these nuclei. 

\begin{center}

\begin{table}[t]
\caption{ Charge root mean square radii $r_c$ [fm] and 
quadrupole deformations $\beta$ obtained with the forces Sk3
and SG2 compared with various works.}

\bigskip

\begin{tabular}{rcccccccc}
\noalign{\smallskip} \\
   & $^{64}$Ge & $^{68}$Se & $^{72}$Kr & $^{76}$Sr & $^{80}$Zr &
$^{84}$Mo & $^{88}$Ru & $^{92}$Pd   \\ 
\noalign{\smallskip}\hline\noalign{\smallskip}
$r_c \quad $ this work Sk3
& 4.034 & 4.127 & 4.216 & 4.344 & 4.406 & 4.343 & 4.397 & 4.458 \\
this work SG2   
& 3.998 & 4.097 & 4.172 & 4.301 & 4.371 & 4.307 & 4.359 & 4.414 \\
   Ref.\cite{ring} 
& 3.985 & 4.089 & 4.180 & 4.283 & 4.350 & 4.333 & 4.370 & 4.410 \\ \\
$\beta   \quad $ this work Sk3  
& 0.199 &-0.216 &-0.268 & 0.408 & 0.431 & 0.005 & 0.019 & 0.111 \\
this work SG2
& -0.192 & -0.219 & -0.246 & 0.416 & 0.002 & 0.001 & 0.038 & 0.103 \\
      Ref.\cite{ring} & 0.217 &-0.285 &-0.358 & 0.410 & 0.437 &-0.247 & 
                        0.107 & 0.112 \\
   Ref.\cite{moeller} & 0.219 & 0.240 &-0.349 & 0.421 & 0.433 & 0.053 & 
                        0.053 & 0.053 \\
\noalign{\smallskip}
\end{tabular}
\end{table}
\end{center}

We can see in Table 1 the microscopically calculated charge root mean 
square radii $r_c$ and quadrupole deformations as defined in 
Eq. (\ref{betadef}). 
Other bulk properties, such as $Q_{EC}$ values, of nuclei in this mass 
region obtained within this formalism have been already published \cite{sarri2},
and they are in a very reasonable agreement with experiment. In this work
we use the experimental $Q_{EC}$ values \cite{audi} in the calculations
of the half-lives.
Since the experimental information on these nuclei is still very little,
we compare in Table 1 our results with the results from 
relativistic mean field calculations of Ref. \cite{ring} and with 
results from systematic calculations \cite{moeller} based on 
macroscopic-microscopic models (finite range droplet macroscopic model 
and folded Yukawa single particle microscopic model). 
The charge radii are quite similar but in most cases the results from SG2
are closer to the relativistic results than those calculated with Sk3.
The quadrupole deformations in Table 1 correspond to the absolute minima
of the energy surfaces. However, we obtain in many cases a second minimum
that in some instances are very close in energy to the absolute one, as it 
can be seen in Fig. 1. In general we observe a good agreement among the 
various theoretical calculations and in those cases where a disagreement
is found, the second minimum mentioned above provides the explanation.
This is the case for example in $^{64}$Ge, where the oblate solution in SG2
($\beta =-0.192$) is accompanied with a prolate solution ($\beta\sim 0.2$)
very close in energy (see Fig. 1). This is also the case of the spherical
solution obtained for $^{80}$Zr with SG2. For $^{80}$Zr with SG2 we also
have a prolate minimum at higher energy. This prolate shape is in agreement
with the ground state solution of Sk3 and with the results from 
Refs. \cite{moeller,ring}. The case of $^{80}$Zr has also been 
studied in Ref. \cite{reinhard}, where the predictions of various Skyrme forces 
were explored. It was shown that most of these forces predict a prolate ground 
state with a deformation compatible with the experimental value ($\beta =0.39$)
\cite{lister}, extracted by relating the measured energy of the first $2^+$ 
excited state with the quadrupole deformation.

The results for GT strength distributions obtained from this formalism have
been already tested against the experimental information available. This 
comparison has been done \cite{feni} in the iron mass region, where charge 
exchange reactions $(np)$ and $(pn)$ have been measured, and $B(GT^+)$ and 
$B(GT^-)$ have been extracted. The agreement with experiment is very good and 
comparable to the results obtained from full shell model 
calculations \cite{caurier}. Results for odd-A nuclei have been analyzed and
compared to experiment in Ref. \cite{sarri3} on the example of Kr isotopes.
The method has also been used to analyze the 
recently measured \cite{isolde} GT strength distributions in $^{74}$Kr and 
$^{76}$Sr. The dependence on deformation of these distributions allows one 
to conclude that the decay pattern of $^{76}$Sr is compatible with a prolate 
shape while that of $^{74}$Kr requires an admixture of oblate and prolate 
shapes.

\begin{center}
\begin{figure}
\resizebox{0.5\textwidth}{!}{%
  \includegraphics{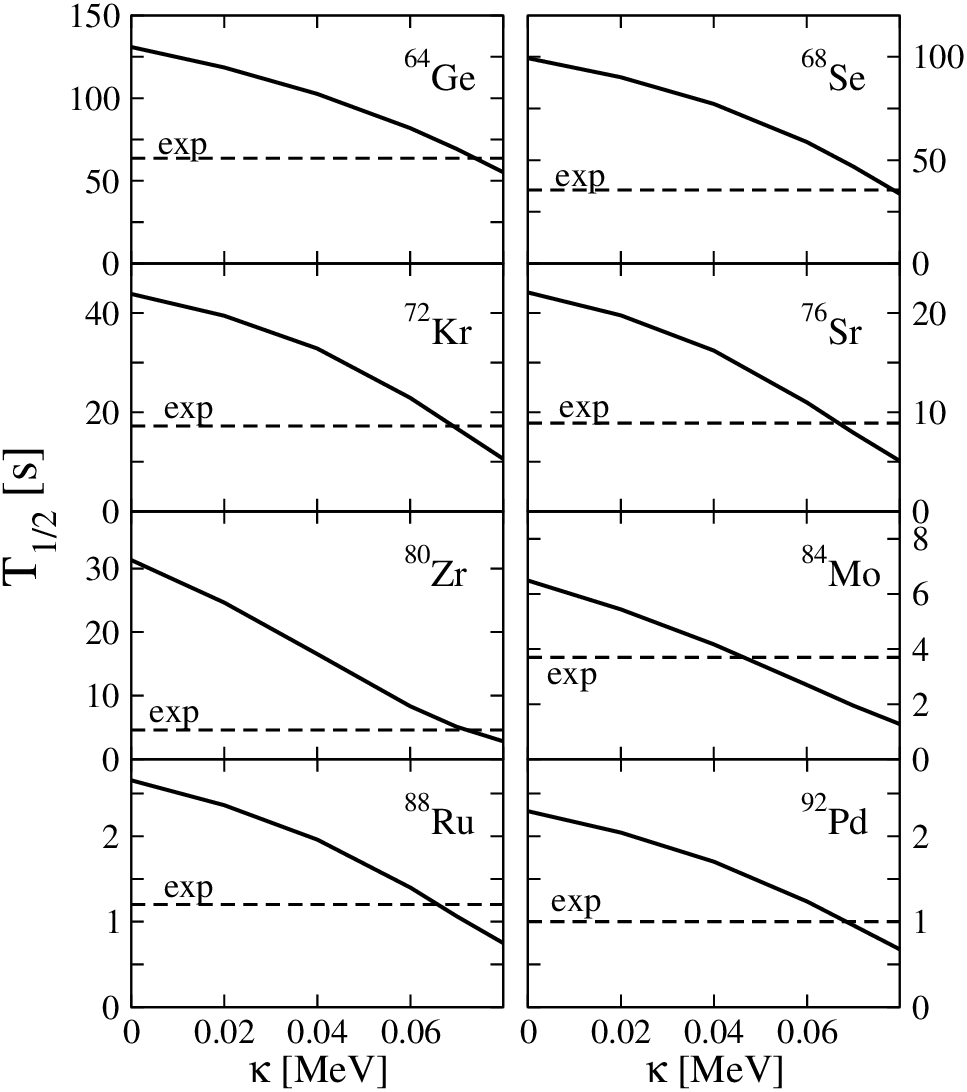}
}
\vskip 0.5cm
\caption{Half-lives of the $N=Z$ waiting points (A=64-92) obtained with the 
force SG2 as a function of 
the $pp$ coupling strength $\kappa$ compared to experiment (dashed lines).}
\label{fig:2}   
\end{figure}
\end{center}

Once the reliability of the method has been contrasted, we proceed now 
with the calculation of the half-lives in the waiting point nuclei. 
Fig. 2 shows the dependence of $T_{1/2}$ on the strength of the $pp$ residual 
interaction, using the Skyrme force SG2. As we have mentioned, the $ph$ 
strength is fixed by the Skyrme force in a consistent way but the $pp$ 
strength remains undetermined. The half-lives for all the waiting points in 
this region are plotted in Fig. 2 as a function of $\kappa ^{pp}_{GT}$ for 
the equilibrium deformations that minimize the energy (see Table 1). 
As we increase $\kappa ^{pp}_{GT}$, the GT strength is 
reduced and shifted to lower energies. Below the $Q_{EC}$ window there is 
a competition between the global reduction of the GT strength and the 
accumulation of extra strength at lower energies. The latter effect
becomes dominant and the half-lives are suppressed as we increase 
$\kappa ^{pp}_{GT}$. The experimental half-life is reproduced in almost all 
cases with values of $\kappa ^{pp}_{GT}$ around 0.07 and therefore this is 
the value postulated to be used in those cases where the half-lives have 
not been measured yet.

In Fig.3 we study the sensitivity of $T_{1/2}$ to deformation. We use the force 
SG2 and  $\kappa ^{pp}_{GT} = 0.07$ MeV.  We can see how 
the HF+BCS half-lives are systematically lower than the QRPA and experiment, 
sometimes even by one order of magnitude. QRPA correlations reduce the mean 
field GT strength and as a result the half-lives increase accordingly. 
Deformation can also change the half-lives by factors up to four. What is 
remarkable is that the experimental half-lives are systematically better 
reproduced for the selfconsistent deformations that minimize the energy, 
as can be seen by comparing Fig. 3 with the SG2 minima in Fig. 1. 
The same feature is found with the Sk3 force. This shows to what extent
selfconsistency plays an important role in the calculation of the 
half-lives. In this respect, it is interesting to note that with the SG2
force, the experimental value of $T_{1/2}$ for $^{80}$Zr is reproduced
at the SG2 equilibrium deformation ($\beta \sim 0$), but not at the 
experimental deformation ($\beta \sim 0.4$) that corresponds to a higher 
local minimum.

\begin{center}
\begin{figure}
\resizebox{0.6\textwidth}{!}{%
  \includegraphics{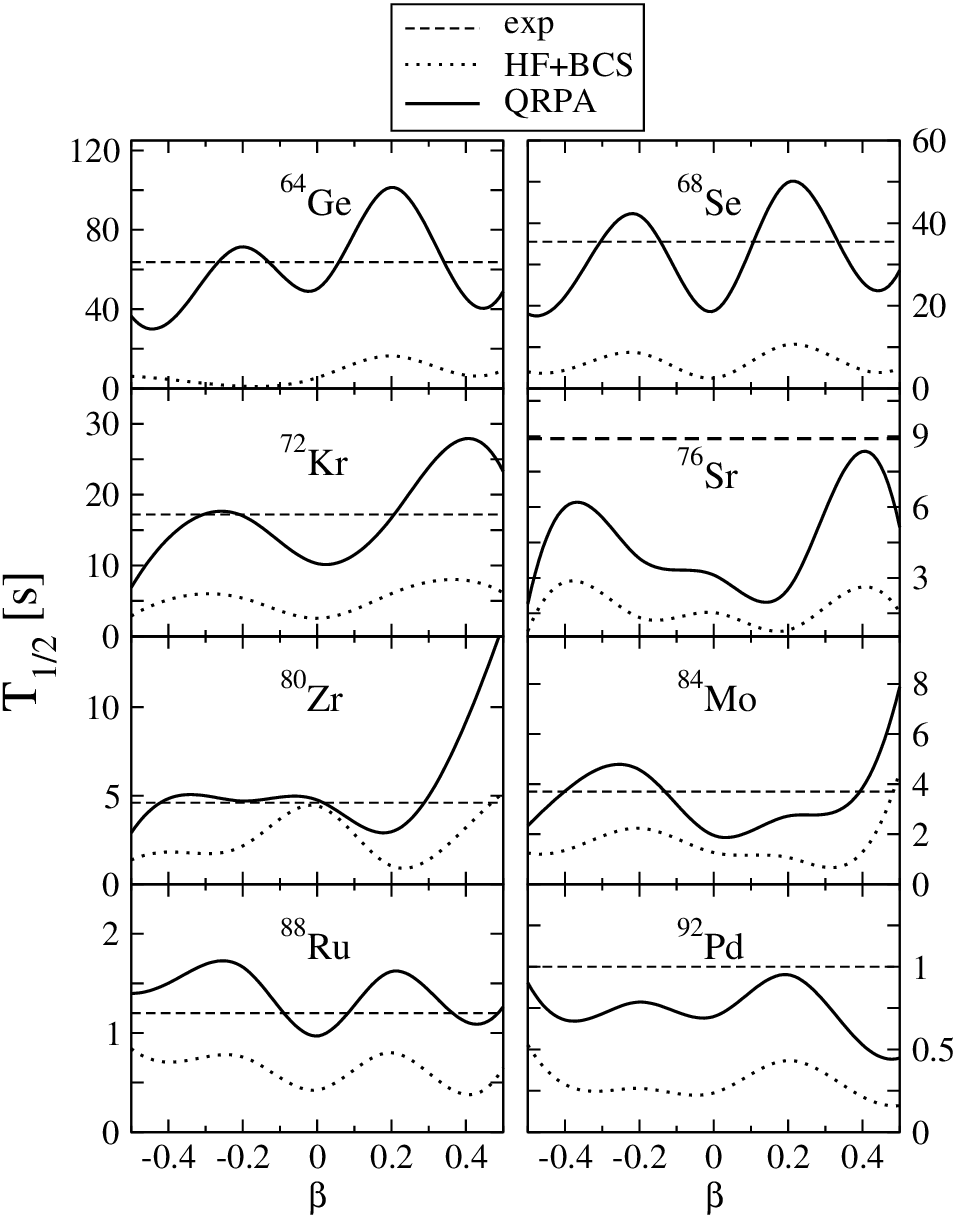}
}
\vskip 0.5cm
\caption{Dependence of the half-life with deformation, using the force SG2. 
The results correspond to HF+BCS (dotted) and QRPA (solid) calculations 
with $\kappa ^{pp}_{GT} = 0.07$ MeV. 
Experimental values are shown by dashed lines.}
\label{fig:3}     
\end{figure}

\begin{figure}
\resizebox{0.7\textwidth}{!}{%
  \includegraphics{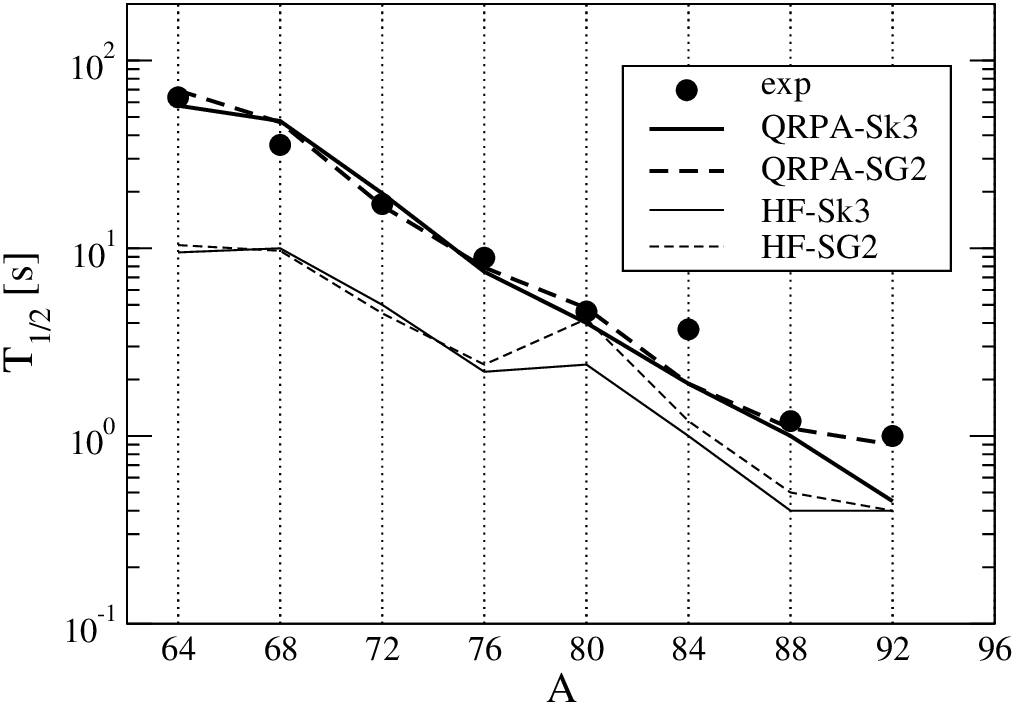}
}
\vskip 0.5cm
\caption{Half-lives of the $N=Z$ waiting points (A=64-92) obtained within
various approaches and compared to experiment \protect\cite{audi} for
A=64,68,72,76,80 and \protect\cite{t12extra} for A=84,88,92.}
\label{fig:4}    
\end{figure}
\end{center}

Finally in Fig. 4 we show the results for the half-lives obtained from our 
approach in the even-even $N=Z$ waiting points from A=64 up to A=92.
The results correspond to the forces Sk3 and SG2 with the deformations 
and coupling strengths residual forces discussed above. 
Again, the HF+BCS results appear systematically below the QRPA 
results and below the experiment. The QRPA results from both forces agree 
nicely with experiment. We get also in general agreement with the QRPA 
results obtained from a different formalism based on Yukawa 
potentials \cite{moellernk}.

\section{Conclusions}
\label{sec:4}

Using a deformed QRPA formalism, based on a selfconsistent quasiparticle
mean field, which includes $ph$ and $pp$ separable residual interactions, 
we have studied the $\beta$-decay properties of several $N=Z$ waiting 
point nuclei involved in the astrophysical rp-process.

We have analyzed the half-lives as a function of deformation and residual
interactions and have found that best agreement with the laboratory 
experimental half-lives is obtained using: a) the selfconsistent 
deformations obtained from the minimization of the energy, and b) residual 
interactions with a consistently derived $ph$ strength and 
$\kappa ^{pp}_{GT}=0.07$ MeV.

The results obtained indicate that this formalism is a useful method for 
reliable calculations of half-lives. This is especially interesting for
applications to: i) cases where no experimental information is available; 
ii) nuclei under different conditions of densities and temperatures; and 
iii) nuclei that are beyond the capability of full shell model 
calculations.
Work under these lines is in progress.

\vskip 1cm

\begin{center}
{\Large \bf Acknowledgments} 
\end{center}
This work was supported by Ministerio de Educaci\'on y Ciencia (Spain) under 
contract number BFM2002-03562. One of us (R.A.-R.) thanks Ministerio de 
Educaci\'on y Ciencia (Spain) for financial support.

\end{document}